\documentclass[aps, prd, twocolumn, nofootinbib,]{revtex4-2}

\usepackage{amsmath}
\usepackage{amsfonts}
\usepackage{wasysym}
\usepackage{graphicx}
\usepackage{comment}
\usepackage{tensor}

\usepackage[colorlinks = true,
            linkcolor = blue,            
            citecolor = blue,
			bookmarks = false]{hyperref}

\def\filetype{pdf}

\def\path{}

\allowdisplaybreaks[1]

\begin{document}



\title{Matter traveling through a wormhole}
\author{Karina Calhoun, Brendan Fay, and Ben Kain}
\affiliation{Department of Physics, College of the Holy Cross, Worcester, Massachusetts 01610, USA}

\begin{abstract}
\noindent We revisit the numerical evolution of Ellis-Bronnikov-Morris-Thorne wormholes, which are constructed with a massless real ghost scalar field.  For our simulations, we have developed a new code based on the standard $3+1$ foliation of spacetime.  We confirm that, for the massless symmetric wormhole, a pulse of regular scalar field causes the wormhole throat to collapse and form an apparent horizon, while a pulse of ghost scalar field can cause the wormhole throat to expand.  As a new result, we show that it is possible for a pulse of regular matter to travel through the wormhole and then to send a light signal back before the wormhole collapses.  We also evolve pulses of matter traveling through massive asymmetric wormholes, which has not previously been simulated.
\end{abstract} 

\maketitle


\section{Introduction}

Considerable attention has been paid to the study of traversable wormholes ever since Morris and Thorne showed that a wormhole geometry in general relativity is possible with matter that violates the null energy condition \cite{Morris:1988cz, VisserBook, LoboBook}.  A Morris-Thorne wormhole may be realized with a massless real ghost scalar field.  By ``ghost," we mean that the kinetic energy has the opposite sign compared to a ``regular" scalar field.  A static wormhole solution in this system was first discovered by Ellis \cite{Ellis:1973yv}.  The solution has zero mass and is symmetric across the wormhole throat.  The solution was generalized by Bronnikov \cite{Bronnikov:1973fh} to a family of static solutions which are generically massive and asymmetric, with the Ellis solution being a special case.  Recent work on such wormholes includes Refs.\ \cite{Gonzalez:2009hn, Bronnikov:2012ch, Tsukamoto:2016qro, Nandi:2016uzg, Dzhunushaliev:2017syc, Konoplya:2018ala, Cremona:2018wkj, Blazquez-Salcedo:2018ipc, Cremona:2019wiy, Bronnikov:2021ods, Yang:2021diz, Sokoliuk:2022xcf, Blazquez-Salcedo:2022kaw}.

As far as we are aware, all numerical simulations of wormholes \cite{Shinkai:2002gv, Gonzalez:2008xk, Doroshkevich:2008xm, Shinkai:2015aqa, Shinkai:2017xkx, Novikov:2019rus, Novikov:2020xwr} have focused on the Ellis-Bronnikov-Morris-Thorne solutions.  Shinkai and Hayward simulated pulses of regular and ghost scalar fields traveling through the massless symmetric wormhole \cite{Shinkai:2002gv}.  Their code was based upon a dual-null formalism \cite{hayward}.  Gonz\'alez et al.\ studied both the massless symmetric and massive asymmetric wormholes \cite{Gonzalez:2008xk}.  They did not simulate a field traveling through the wormhole, but instead considered a perturbation applied directly to one of the metric fields.  The numerical formalism of Gonz\'alez et al.\ bears similarities to the $3+1$ formalism we use here, but they did not incorporate the extrinsic curvature.  Doroshkevich et al.\ \cite{Doroshkevich:2008xm} extended the work of \cite{Shinkai:2002gv} and, like \cite{Shinkai:2002gv}, focused on the massless symmetric wormhole and used a dual-null formalism, as did subsequent studies \cite{Novikov:2019rus, Novikov:2020xwr}.  Shinkai and Torii used the dual-null formalism to study higher-dimensional versions of the massless symmetric wormhole in the context of Einstein-Gauss-Bonnet gravity \cite{Shinkai:2015aqa, Shinkai:2017xkx}.

In this work, we numerically simulate regular and ghost scalar fields traveling through both the massless symmetric and massive asymmetric wormholes.  For our simulations, we have developed a new code based on the standard $3+1$ foliation of spacetime \cite{AlcubierreBook, BaumgarteBook}.  Using our code, we confirm results found in \cite{Shinkai:2002gv}.  In particular, we confirm that a pulse of regular scalar field traveling through the massless symmetric wormhole causes the throat to collapse and form an apparent horizon, as does a pulse of ghost scalar field with negative amplitude, while a pulse of ghost scalar field with positive amplitude causes the throat to expand.  As a new result, we show that it is possible for a pulse of regular field to travel through the wormhole and then to send a light signal back before the wormhole collapses.  We also study matter traveling through massive asymmetric wormholes, which has not previously been simulated.  Such wormholes are not stationary, but move.  We find again that a pulse of regular scalar field causes the wormhole throat to collapse and form an apparent horizon and a pulse of ghost scalar field can cause the wormhole throat to expand.

In the next section, we briefly describe the $3+1$ formalism and present the equations we will be using.  These include equations for both the metric and matter fields.  In Sec.\ \ref{sec:ID}, we consider the static limit of these equations and review the Ellis-Bronnikov-Morris-Thorne solutions for static wormholes.  We then describe adding a pulse of regular or ghost scalar field and how we use these solutions as initial data for our simulations.  In Sec.\ \ref{sec:numerics}, we briefly describe aspects of our numerical code.  We present our results for the massless symmetric wormhole in Sec.\ \ref{sec:massless results} and for massive asymmetric wormholes in Sec.\ \ref{sec:massive results}.  We conclude in Sec.\ \ref{sec:conclusion}.  In an appendix, we present tests of our code.


\section{Equations}
\label{sec:equations}

The general spherically symmetric metric can be written \cite{AlcubierreBook, BaumgarteBook}
\begin{equation}
\begin{split}
ds^2 &= -\left(\alpha^2 - A \beta^{r\,2}\right) dt^2 + 2 A \beta^r dt dr
 + dl^2
\\
dl^2 &=  A dr^2 + C \left( d\theta^2 + \sin^2\theta d\phi^2 \right).
\end{split}
\end{equation}
In the $3+1$ formalism, we foliate spacetime into a continuum of time slices, where each time slice is a spatial hypersurface.  $dl^2$ is the spatial metric on an individual time slice.  $A$ and $C$ parametrize the spatial metric, where $A$ tells us about the physical distance between coordinates and $C$ is the squared areal radius.  We define
\begin{equation}
R \equiv \sqrt{C}
\end{equation}
as the areal radius, so that the area of a two-sphere is $4\pi R^2$.  $\alpha$ is the lapse and $\beta^r$ is the only nonzero component of the shift vector.  As we move from one time slice to the next, $\alpha$ tells us the rate at which proper time increases and $\beta^r$ tells us how the coordinates shift.  Both $\alpha$ and $\beta^r$ are gauge functions, in that once initial data are loaded onto the initial time slice $\alpha$ and $\beta^r$ can, in principle, be chosen arbitrarily.  In addition to these fields, there is the intrinsic curvature, $K\indices{^i_j}$, which describes how a time slice sits in the larger spacetime.  The only nonzero and independent components of the extrinsic curvature are $K\indices{^r_r}$ and $K_T \equiv 2K\indices{^\theta_\theta} = 2K\indices{^\phi_\phi}$.  All quantities are functions of $t$ and $r$.

In a wormhole geometry, $r$ is not restricted to be non-negative, but instead takes values $-\infty < r < \infty$.  If $R$ is nonzero at its minimum, then there exists a wormhole throat which connects the two sides of the minimum.  We define the wormhole throat radius to be 
\begin{equation}
R_\text{th}(t) \equiv R(t,r_\text{min}),
\end{equation}
where, on a given time slice, $R$ has its minimum at $r_\text{min}$.

We use units such that $c = G = \hbar = 1$.  The Einstein field equations, $G\indices{^\mu_\nu} = 8\pi T\indices{^\mu_\nu}$, where $G\indices{^\mu_\nu}$ is the Einstein tensor and $T\indices{^\mu_\nu}$ is the energy-momentum tensor, in the spherically symmetric $3+1$ formalism lead to the evolution equations
\begin{widetext}
\begin{equation} \label{evo eqs} 
\begin{split}
\partial_t A &=   - 2\alpha A K\indices{^r_r}
+ \beta^r \partial_r A + 2 A \partial_r \beta^r
\\
\partial_t C &= -\alpha C  K_T + \beta^r \partial_r C
\\
\partial_t K\indices{^r_r} 
&=
\alpha \left[ \frac{(\partial_r C)^2}{4AC^2}
- \frac{1}{C} 
+ (K\indices{^r_r})^2 
- \frac{1}{4}K_T^2 
+ 4\pi (S +  \rho - 2 S\indices{^r_r} )
\right]
- \frac{\partial_r^2 \alpha}{A} + \frac{(\partial_r \alpha)(\partial_r A)}{2A^2}
+ \beta^r \partial_r K\indices{^r_r}
\\
\partial_t K_T
&= \alpha \left[ \frac{1}{C} 
- \frac{(\partial_r C)^2 }{4AC^2}
+ \frac{3}{4} (K_T)^2 
+ 8\pi S\indices{^r_r}
\right]
- \frac{(\partial_r \alpha)(\partial_r C)}{AC}
+ \beta^r \left[
\frac{\partial_r C}{2C} \left(2 K\indices{^r_r} -  K_T \right) 
-8\pi S_r
\right]
\end{split}
\end{equation}
and the Hamiltonian and momentum constraint equations,
\begin{equation} \label{con eqs}
\begin{split}
\partial_r^2 C &=
A + \frac{(\partial_r A)(\partial_r C)}{2A} 
+ \frac{(\partial_r C)^2}{4C} 
+ ACK_T \left( K\indices{^r_r} + \frac{1}{4} K_T \right)
- 8\pi AC \rho
\\
\partial_r K_T &=  
\frac{\partial_r C}{2C} \left(2 K\indices{^r_r} -  K_T \right)  -8\pi S_r,
\end{split}
\end{equation}
\end{widetext}
where
\begin{equation} \label{matter functions defs}
\begin{split}
\rho &= -T\indices{^t_t} + \beta^r T\indices{^t_r}
\\
S\indices{^r_r} &= \frac{1}{A} T_{rr}
\\
S\indices{^\theta_\theta} &= \frac{1}{C} T_{\theta\theta}
\\
S_r &= \alpha T\indices{^t_r}
\end{split}
\end{equation}
and $S = S\indices{^r_r} + 2 S\indices{^\theta_\theta}$ depend on the energy-momentum tensor and parametrize the matter sector.

Ellis-Bronnikov-Morris-Thorne wormholes are constructed with a massless real ghost scalar field.  As mentioned in the Introduction, ``ghost" means that the kinetic energy has the opposite sign compared to the kinetic energy of a ``regular" scalar field.  Since we are interested in sending a regular field through the wormhole, we will include both ghost and regular scalar fields.  The Lagrangian for our matter sector is then
\begin{equation}
\mathcal{L} = - \frac{\eta_\phi}{2} \partial_\mu \phi \partial^\mu \phi
- \frac{\eta_\chi}{2} \partial_\mu \chi \partial^\mu \chi.
\end{equation}
$\phi$ is the ghost field that forms the wormhole and $\chi$ is the regular field and thus
\begin{equation}
\eta_\phi = -1,
\qquad
\eta_\chi = +1.
\end{equation}
We minimally couple the Lagrangian to gravity through $\mathcal{L} \rightarrow \sqrt{-\det(g_{\mu\nu})} \, \mathcal{L}$, from which the equations of motion and energy-momentum tensor follow straightforwardly.  To write the equations of motion, we define the auxiliary fields
\begin{equation}
\begin{aligned}
\Phi_\phi &\equiv \partial_r \phi&\quad
\Pi_\phi &\equiv \frac{C \sqrt{A}}{\alpha} \left(\partial_t \phi - \beta^r \partial_r \phi \right)
\\
\Phi_\chi &\equiv \partial_r \chi& \quad
\Pi_\chi &\equiv \frac{C \sqrt{A}}{\alpha} \left(\partial_t \chi - \beta^r \partial_r \chi \right)
\end{aligned}
\end{equation}
and the equations of motion are
\begin{equation} \label{ghost eom}
\begin{split}
\partial_t \phi &= \frac{\alpha }{ C\sqrt{A}}\Pi_\phi + \beta^r \Phi_\phi
\\
\partial_t \Phi_\phi &= \partial_r \left( \frac{\alpha}{C \sqrt{A}} \Pi_\phi + \beta^r \Phi_\phi \right)
\\
\partial_t \Pi_\phi &= 
\partial_r \left( \frac{\alpha C}{\sqrt{A}} \Phi_\phi + \beta^r \Pi_\phi \right)
\end{split}
\end{equation}
and 
\begin{equation} \label{reg eom}
\begin{split}
\partial_t \chi &= \frac{\alpha }{ C\sqrt{A}}\Pi_\chi + \beta^r \Phi_\chi
\\
\partial_t \Phi_\chi &= \partial_r \left( \frac{\alpha}{C \sqrt{A}}\Pi_\chi  + \beta^r \Phi_\chi \right)
\\
\partial_t \Pi_\chi &= 
\partial_r \left( \frac{\alpha C}{\sqrt{A}} \Phi_\chi + \beta^r \Pi_\chi \right).
\end{split}
\end{equation}
From the energy-momentum tensor, we have the matter functions
\begin{equation} \label{matter functions}
\begin{split}
\rho &=  \frac{1}{2A} \left[\eta_\phi \left( \frac{\Pi^2_\phi}{C^2} + \Phi^2_\phi\right) + \eta_\chi \left( \frac{\Pi^2_\chi}{C^2} + \Phi^2_\chi \right)\right]
\\
S\indices{^r_r} &=  \frac{1}{2A} \left[\eta_\phi \left( \frac{\Pi^2_\phi}{C^2} + \Phi^2_\phi\right) + \eta_\chi \left( \frac{\Pi^2_\chi}{C^2} + \Phi^2_\chi \right)\right]
\\
S\indices{^\theta_\theta} &=  \frac{1}{2A} \left[\eta_\phi \left( \frac{\Pi^2_\phi}{C^2} - \Phi^2_\phi\right) + \eta_\chi \left( \frac{\Pi^2_\chi}{C^2} - \Phi^2_\chi \right)\right]
\\
S_r &= - \frac{1}{C\sqrt{A}} \left( \eta_\phi \Pi_\phi \Phi_\phi + \eta_\chi \Pi_\chi \Phi_\chi \right),
\end{split}
\end{equation}
which follow from Eq.\ (\ref{matter functions defs}).

The solution to the equations listed in this section give the numerical evolution of the system.  Note that the fields $\phi$ and $\chi$ decouple and do not have to be determined.  Instead, for matter fields, we need only determine $\Phi_\phi$, $\Pi_\phi$, $\Phi_\chi$, and $\Pi_\chi$.

We can use the solution to compute various quantities.  For example, the Misner-Sharp mass function, $m(t,r)$, in spherically symmetric systems is given by \cite{AlcubierreBook, BaumgarteBook}
\begin{equation}\label{m function}
m 
= \frac{1}{8 \sqrt{C}}
\left[ 4C 
+ (CK_T )^2 - \frac{(\partial_r C)^2}{A}  \right].
\end{equation}
The ADM mass, $M$, is given by the large $r$ limit of $m$.  In a wormhole geometry, we can take $r\rightarrow \pm \infty$, and we should not expect the two limits to give the same value.  Instead, the two limits give the total mass as viewed from either side of the wormhole.

We will be interested in whether an apparent horizon forms, which we will use as an indicator for a black hole.  Apparent horizons satisfy \cite{AlcubierreBook, BaumgarteBook, Gonzalez:2008xk}
\begin{equation} \label{apparent horizon}
C \sqrt{A}  K_T \mp \partial_r C = 0,
\end{equation}
where we use the upper sign for $r > r_\text{min}$, where $r_\text{min}$ marks the minimum of $C$, and the lower sign for $r < r_\text{min}$.

Finally, we can compute null geodesics, $r_\text{null}(t)$, which obey
\begin{equation} \label{null def}
\frac{dr_\text{null}}{dt} = \pm \frac{\alpha}{\sqrt{A}} - \beta^r,
\end{equation}
where we choose the sign depending on whether we want to compute a right-moving or left-moving geodesic.


\section{Initial data}
\label{sec:ID}

To numerically evolve the system, we must place initial data on the initial time slice.  Our code can then evolve the initial data forward in time.  Our initial data will include a wormhole and a pulse of ghost or regular matter.

We assume our initial data are time-symmetric, which sets $\beta^r = K\indices{^i_j} = 0$ \cite{BaumgarteBook}.  Under this assumption, the Hamiltonian constraint in Eq.\ (\ref{con eqs}) and the $K_T$ evolution equation in Eq.\ (\ref{evo eqs}) become
\begin{align}  \label{static eqs}
\partial_r^2 C &=
A + \frac{(\partial_r A)(\partial_r C)}{2A} 
+ \frac{(\partial_r C)^2}{4C}
- 8\pi AC \rho
\notag
\\
0
&= \frac{1}{C} - \frac{(\partial_r \alpha)(\partial_r C)}{\alpha AC}  
- \frac{(\partial_r C)^2 }{4AC^2} 
+ 8\pi  S\indices{^r_r},
\end{align}
the evolution equations for $K\indices{^r_r}$ and $K_T$ in Eq.\ (\ref{evo eqs}) can be combined to give
\begin{equation} \label{static alpha eq}
\begin{split}
\partial^2_r \alpha& = 
\left(
\frac{\partial_r A}{2A}
- \frac{\partial_r C}{ C}
\right) \partial_r \alpha
+ 4\pi\alpha A(\rho + S),
\end{split}
\end{equation}
and the Misner-Sharp mass function in (\ref{m function}) reduces to
\begin{equation}\label{static m function}
m =  \frac{\sqrt{C}}{2}
\left[1 
 - \frac{ (\partial_r C)^2}{4 A C}  \right].
\end{equation}


\subsection{Static wormholes}

We first discuss the wormhole before including pulses.  For now, then, we drop the regular field by setting $\chi = 0$.  We use a static wormhole solution for our initial data.  By static, we mean that spacetime is time independent, which can be achieved by assuming $\phi$ is time independent.  The ghost field equations of motion in (\ref{ghost eom}) reduce to
\begin{equation} \label{static eom}
0 = \partial_r \left( \frac{\alpha C}{\sqrt{A}} \Phi_\phi \right)
\end{equation}
and the matter functions in (\ref{matter functions}) reduce to
\begin{equation} \label{static matter functions}
\rho = S\indices{^r_r} = - S\indices{^\theta_\theta} = \eta_\phi \frac{\Phi_\phi^2}{2A},
\end{equation}
along with $S_r = 0$.

Static wormhole solutions can be found analytically \cite{Ellis:1973yv, Bronnikov:1973fh, Armendariz-Picon:2002gjc, Gonzalez:2008wd} for
\begin{equation} \label{static analytical wormhole}
\alpha = \frac{1}{\sqrt{A}}, \qquad
C = A(r_0^2 + r^2),
\end{equation}
where $r_0$ is a constant.  For $\alpha = 1/\sqrt{A}$, the top equation in (\ref{static eqs}) is unchanged, while the bottom equation and Eq.\ (\ref{static alpha eq}) become
\begin{equation} \label{alpha gauge}
\begin{split}
\partial_r^2 A 
&= \frac{2 (\partial_r A)^2}{A}
- \frac{(\partial_r A)(\partial_r C)}{C}
- 8\pi A^2(\rho + S)
\\
0 &=  \frac{(\partial_r A)(\partial_r C)}{2A^2C}
+ \frac{1}{C} 
- \frac{(\partial_r C)^2}{4AC^2}  
+ 8\pi  S\indices{^r_r}.
\end{split}
\end{equation}
A look at the matter functions in (\ref{static matter functions}) shows that $\rho = -S$, where $S = S\indices{^r_r} + 2S\indices{^\theta_\theta}$, so that the matter functions cancel out in the top equation in (\ref{alpha gauge}).  Plugging in $C = A(r_0^2 + r^2)$, the result can be solved for $A$:
\begin{equation} \label{static A}
A(r) = A_2 \exp \left[ - 2 A_1 \tan^{-1} (r/r_0) \right],
\end{equation}
where $A_1$ and $A_2$ are arbitrary integration constants.  $A$ cancels out in Eq.\ (\ref{static eom}) after using (\ref{static analytical wormhole}) and the result can be solved for $\Phi_\phi$:
\begin{equation} \label{static Phi}
\Phi_\phi = \phi_1\frac{r_0}{r_0^2 + r^2},
\end{equation}
which can be integrated to give $\phi(r) = \phi_1 \tan^{-1} (r/r_0) + \phi_2$,
where $\phi_1$ and $\phi_2$ are arbitrary integration constants.

We now fix some of the integration constants.  For our static wormhole solutions, the metric is given by
\begin{equation}
ds^2 = -\frac{1}{A} dt^2 + A dr^2 + A(r_0^2+r^2) (d\theta^2 + \sin^2\theta d \phi^2).
\end{equation}
We require spacetime to be asymptotically flat.  A look at the metric suggests that this can be accomplished by requiring $A\rightarrow 1$ in the large $r$ limit.  However, from Eq.\ (\ref{static A}) we have
\begin{equation}
\lim_{r\rightarrow \pm \infty} A(r) =  A_2 e^{\mp A_1 \pi },
\end{equation}
which indicates that, unless $A_1 = 0$, we can have $A \rightarrow 1$ on only one side of the wormhole at a time.  A simple rescaling,
\begin{equation}
t \rightarrow \sqrt{\lambda} \, t, \qquad
r \rightarrow  \frac{r}{\sqrt{\lambda}}, \qquad
r_0 \rightarrow \frac{r_0}{\sqrt{\lambda}}, \qquad
A \rightarrow \lambda  A ,
\end{equation}
where $\lambda$ is a positive constant, leaves the metric unchanged.  For $\lambda = e^{\pm A_1 \pi}/A_2$, the $\pm$ side has $A\rightarrow 1$, which shows that, in fact, both sides of the wormhole are asymptotically flat for any non-negative value for  $A_2$.  Following Ref.\ \cite{Gonzalez:2008xk}, we choose 
\begin{equation} \label{A2}
A_2 = 1.
\end{equation}
Plugging our results into the top equation in (\ref{static eqs}) leads to
\begin{equation} \label{phi1}
\phi_1 = \sqrt{- \frac{1 + A_1^2}{4\pi \eta_\phi}}.
\end{equation}
In addition to determining $\phi_1$, this equation explains why it is a ghost field, with $\eta_\phi = -1$, that is required to find a wormhole solution.

It is now a simple matter to show that the areal radius, $R=\sqrt{C}$, has a minimum at
\begin{equation} \label{static rmin}
r_\text{min} = A_1 r_0
\end{equation} 
and that the wormhole throat radius is given by
\begin{equation} \label{static Rth}
R_\text{th} = R(r_\text{min}) = r_0 \sqrt{A_2(1 + A_1^2)} \, e^{- A_1 \tan^{-1}(A_1)}.
\end{equation}
For positive $A_2$, $R_\text{th} > 0$ and we have a wormhole solution.

We have constructed a family of analytical static wormhole solutions.  The solutions are identified by the constant $A_1$, which immediately determines the constant $\phi_1$ from Eq.\ (\ref{phi1}).  These constants then determine $\alpha$, $A$, $C$, and $\Phi_\phi$ from Eqs.\ (\ref{static analytical wormhole}), (\ref{static A}), and (\ref{static Phi}).  We note that it is straightforward to show that these solutions satisfy the bottom equation in (\ref{alpha gauge}), as required.  Using the solutions in the Misner-Sharp mass function in (\ref{static m function}) and then taking the large $r$ limit, we find the total mass on both sides of the wormhole,
\begin{equation}
\begin{split}
M_\pm = \lim_{r\rightarrow \pm \infty} m(r) &= \pm A_1 r_0 \left( A_2 e^{\mp A_1 \pi} \right)^{1/2}
\\
&= \pm A_1 r_0 e^{\mp A_1 \pi/2},
\end{split}
\end{equation}
where we used (\ref{A2}) in the bottom line.  For $A_1 = 0$, we have the massless symmetric wormhole.  Nonzero values of $A_1$ describe massive asymmetric wormholes.


\subsection{Regular and ghost pulses}

We will numerically simulate a pulse of regular or ghost scalar field traveling though the wormhole, using a Gaussian shape for the pulse.  For the regular field, we use
\begin{equation} \label{chi pulse}
\Phi_\chi(0, r) = \varepsilon_\chi e^{-(r - r_\chi)^2/\sigma_\chi^2},
\end{equation}
where $\varepsilon_\chi$, $r_\chi$, and $\sigma_\chi$ are constants that describe the amplitude, center, and spread of the pulse.  Further, we assume the pulse is initially at rest, which sets $\Pi_\chi(0,r) = 0$.  This pulse represents a spherical shell of regular massless scalar field.  As the pulse begins stationary, it will split into an ingoing piece that heads toward the wormhole and an outgoing piece that heads away from the wormhole.  Of course, our focus is with the ingoing piece.  

For a pulse of massless ghost scalar field, we use
\begin{equation} \label{ghost pulse ID}
\Phi_\phi(0,r) = \phi_1 \frac{r_0}{r_0^2 + r^2} 
+ \varepsilon_\phi e^{-(r - r_\phi)^2/\sigma_\phi^2},
\end{equation}
where the first term represents the wormhole, as described in the previous subsection, and the second term is analogous to Eq.\ (\ref{chi pulse}).  We take $\Pi_\phi(0,r) = 0$.

The metric functions $A$ and $C$ must be consistent with the matter functions $\Phi_\phi$, $\Pi_\phi$, $\Phi_\chi$, and $\Pi_\chi$.  We have in mind a matter pulse heading toward and traveling through a previously formed wormhole.  In this sense, the matter pulse can be thought of as a perturbation to the wormhole.  To model this, we assume the region near the wormhole throat, i.e.\ near $r_\text{min}$ where the areal radius $R$ has its minimum, is unaffected by the pulse on the initial time slice.  Since $r_\text{min}$ will be near $r=0$, we keep $A(0)$, $A'(0)$, $C(0)$, and $C'(0)$ equal to their values in the absence of a pulse and, using these values, obtain $A$ and $C$ by numerically integrating the top equations in (\ref{static eqs}) and (\ref{alpha gauge}) outward from $r=0$.

The equations are independent of the sign of $\xi_\chi$, and without loss of generality, we take $\xi_\chi$ to be positive.  This is not the case for $\xi_\phi$, because of the first term in Eq.\ (\ref{ghost pulse ID}).  In Secs.\ \ref{sec:massless results} and \ref{sec:massive results}, we present results using $r_\phi$, $r_\chi = 3.0 r_0$ and $\sigma_\phi$, $\sigma_\chi = 0.25r_0$.  We have found that other values lead to qualitatively similar results.  For amplitudes we use $\xi_\phi$, $\xi_\chi = 0$ or $0.002/ r_0$.  If the amplitude is too large, the pulse can form a black hole.  Aside from this, and as long as $\xi_\phi$ is non-negative, we have found that other values lead to qualitatively similar results.  Although we do not show results for negative values of $\xi_\phi$, we have run such simulations and will discuss them below.


\section{Numerics}
\label{sec:numerics}

Our numerical simulations begin with the placement of initial data for $A$, $C$, $\Phi_\phi$, and $\Phi_\chi$ on the initial time slice.  The remaining fields, except for $\alpha$ and possibly $\beta^r$, are set to zero.  As mentioned, $\alpha$ and $\beta^r$ are gauge fields and can be set arbitrarily.  We choose to keep $\beta^r(t,r) = 0$ throughout the simulation.  For our time slicing condition we primarily use
\begin{equation} \label{A slicing}
\alpha(t,r) = \sqrt{A(t,r)}.
\end{equation}
This slicing condition was shown to work well in Ref.\ \cite{Gonzalez:2008xk}, though it was also shown not to have strong singularity avoidance properties.  We comment on this and briefly consider different slicing conditions in the next two sections.

The next step is to numerically solve the equations listed in Sec.\ \ref{sec:equations}.  To do this, we discretize both space and time.  Our spatial grid has uniform spacing $\Delta r$ and includes $r = 0$ and our temporal grid has uniform time step $\Delta t$.  We have some choice in which equations to solve.  A straightforward possibly is to exclusively use evolution equations and solve Eqs.\ (\ref{evo eqs}), (\ref{ghost eom}), and (\ref{reg eom}).  Alternatively, we could make use of the constraint equations in (\ref{con eqs}).  Typically, the use of constraint equations improves numerical stability.  To use constraint equations, the value of the field must be known somewhere on a time slice.

In our code, we use the evolution equation for $K_T$ in (\ref{evo eqs}) to evolve $K_T$ from time step $n$ to time step $n+1$.  We then choose one value for $K_T$ on time step $n+1$, typically at $r = 0$, and use it to solve the momentum constraint in (\ref{con eqs}) for $K_T$.  The actual $K_T$ values we use at time step $n+1$ are those obtained from the momentum constraint.  We have found that this greatly improves numerical stability.  We could implement something similar for $C$ using the Hamiltonian constraint, but do not do so.

We solve the evolution equations using the method of lines, in which we finite difference spatial derivatives and solve time derivatives using third-order Runge-Kutta.  For those evolution equations that contain spatial derivatives, we use simple outgoing spherical wave boundary conditions
\begin{equation}
(\partial_t \pm \partial_r)(rf) = 0,
\end{equation}
with the upper sign for the positive outer boundary and the lower sign for the negative outer boundary, where $f$ is the field.  For those evolution equations that do not contain spatial derivatives, we solve the evolution equation all the way up to the outer boundary.  We place the outer boundary at $r/r_0 = \pm 100$, which is sufficiently far away that reflections are negligible.

In our simulations, we use time step $\Delta t = 0.5 \Delta r$ and scale
\begin{equation}
t \rightarrow r_0 t, \qquad
r \rightarrow r_0 r, \qquad
C \rightarrow r_0^2 C,
\end{equation} 
along with $r_{\phi,\chi} \rightarrow r_0 r_{\phi,\chi}$, $\sigma_{\phi,\chi} \rightarrow r_0 \sigma_{\phi,\chi}$, and $\xi_{\phi,\chi} \rightarrow \xi_{\phi,\chi}/r_0$ which is equivalent to setting $r_0 = 1$.  In the Appendix we present various code tests and show that our code is second order convergent.  We do this by evolving the system using different grid spacings, $\Delta r$.  For the results presented in the upcoming sections, we use the smallest grid spacing used in the Appendix, $\Delta r = 0.00125$.


\section{Massless symmetric wormhole}
\label{sec:massless results}

\begin{figure*}
\centering
\includegraphics[width=7in]{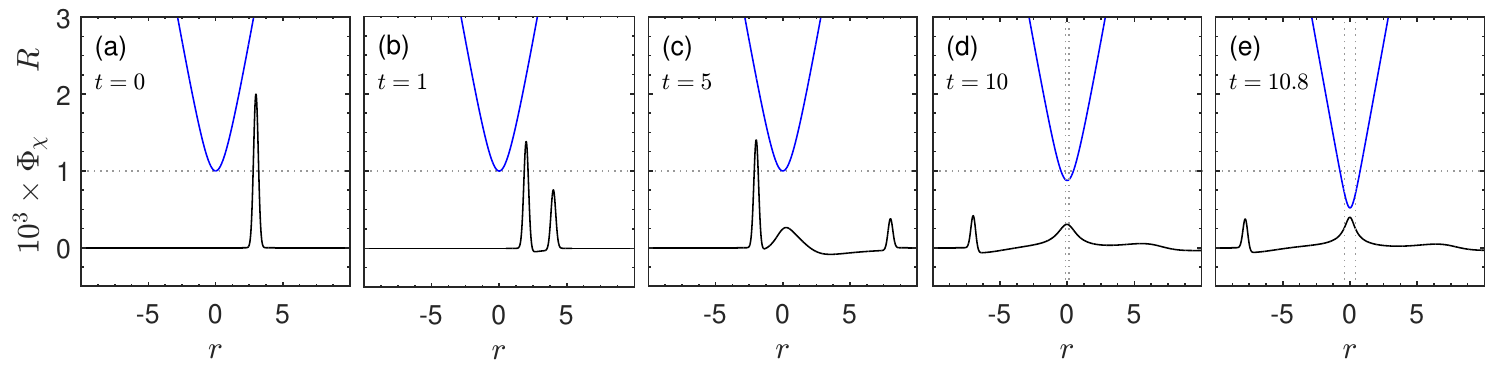}
\caption{Snapshots of the evolution of a pulse of regular scalar field, $\Phi_\chi$ (black curve), and the areal radius, $R$ (blue curve), for the massless symmetric wormhole defined by $A_1 = 0$.  The horizontal dotted line, which is the same in (a)--(e), marks the initial value of the wormhole throat radius and helps us to see if the throat expands or collapses.  The vertical dotted lines in (d) and (e) show apparent horizons.  See the main text for details.}
\label{fig:massless regular}
\end{figure*}

In this section, we present results for the numerical evolution of the massless symmetric wormhole, which is defined by $A_1 = 0$.  In Fig.\ \ref{fig:massless regular}, we show results for the evolution of a pulse of regular scalar field, $\Phi_\chi$ (black curve), and the areal radius, $R$ (blue curve).
The horizontal dotted line, which is the same in each plot, marks the initial value of the wormhole throat, $R_\text{th}$, and helps us to see if the throat expands or collapses.  The pulse begins at $t = 0$ as a Gaussian and then splits into an inward traveling piece and outward traveling piece.  This is expected and is standard for a scalar field in spherical symmetry.  Our focus is on the inward traveling piece, which we can see traveling through the wormhole, i.e traveling from one side of $r_\text{min} = 0$ to the other side, in Figs.\ \ref{fig:massless regular}(b) and \ref{fig:massless regular}(c).  Interestingly, we see in Fig.\ \ref{fig:massless regular}(c) that some matter gets caught in the wormhole.  This matter stays in the wormhole as the piece that traveled through continues traveling leftward away from the wormhole, as seen in Figs.\ \ref{fig:massless regular}(d) and \ref{fig:massless regular}(e).  Starting in Fig.\ \ref{fig:massless regular}(d), we show the location of an apparent horizon with the vertical dotted lines.  The coordinates of the apparent horizon are seen to expand in Fig.\ \ref{fig:massless regular}(e).  Also starting in Fig.\ \ref{fig:massless regular}(d), we see the collapsing of the wormhole, since $R_\text{th}$ decreases, which continues in Fig.\ \ref{fig:massless regular}(e).

\begin{figure}
\centering
\includegraphics[width=3.25in]{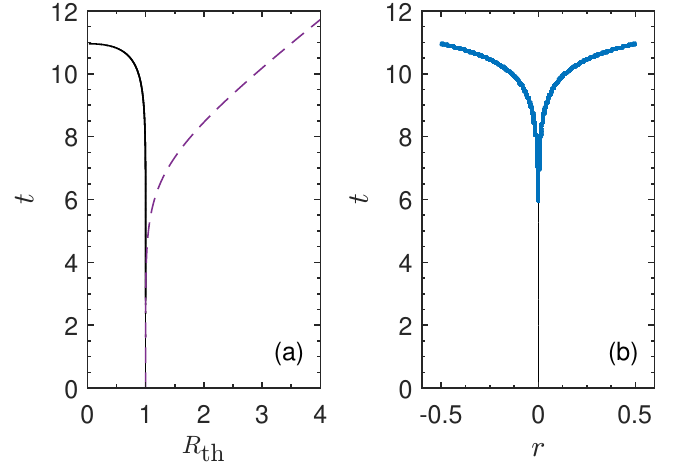}
\caption{(a) The solid black curve plots the wormhole throat radius, $R_\text{th}$, as a function of time for the same evolution shown in Fig.\ \ref{fig:massless regular}.  The pulse of regular matter traveling through the wormhole causes the wormhole throat to collapse.  The dashed purple curve shows the wormhole throat radius for a pulse of ghost matter with positive amplitude, which causes the throat to expand.  (b) The vertical black line marks the location of the wormhole throat for the same evolution shown in Fig.\ \ref{fig:massless regular}.  The blue curve plots the apparent horizon.}
\label{fig:massless regular R}
\end{figure}

For the same evolution shown in Fig.\ \ref{fig:massless regular}, we show in Fig.\ \ref{fig:massless regular R}(a) the wormhole throat radius $R_\text{th}$ as a function of time as the solid black curve.  We can see more clearly how the wormhole throat is collapsing.  In Fig.\ \ref{fig:massless regular R}(b), the vertical line plots $r_\text{min}$, the location of the wormhole throat.  We plot $r_\text{min}$ until the formation of the apparent horizon, which is shown by the blue curve.

We now consider sending a ghost pulse through the wormhole using the initial data in (\ref{ghost pulse ID}).  Ideally, we would present a plot analogous to Fig.\ \ref{fig:massless regular}.  Unfortunately, this is not possible since $\Phi_\phi$ is dominated by the wormhole contribution and we cannot extract out the pulse.  For a ghost pulse with positive amplitude, we find that the wormhole throat expands, as shown by the purple dashed curve in Fig.\ \ref{fig:massless regular R}(a), and that no apparent horizon forms.  On the other hand, for a ghost pulse with negative amplitude, which we do not show, we find that the wormhole throat collapses.  The reason for this is that the negative amplitude shrinks the initial magnitude of $\Phi_\phi$, which increases the initial energy density (since the ghost field has negative energy density), just as a regular pulse does.

In general, when a pulse of regular scalar field is sent through the wormhole, we find that the wormhole collapses and an apparent horizon forms.  When a pulse of ghost field with negative amplitude is sent through, we find the same, while for a pulse of ghost field with positive amplitude we find that the wormhole expands and no apparent horizon forms.  This is true even for relatively weak pulses.  This suggests that the massless symmetric wormhole, which has been shown to be linearly unstable with a single unstable mode \cite{Gonzalez:2008wd}, sits right at the transition between expansion and collapse to a black hole.  

The results presented so far in this section are consistent with those in Ref.\ \cite{Shinkai:2002gv}.  However, we find the plots in Fig.\ \ref{fig:massless regular} to be particularly intuitive and a fascinating depiction of matter traveling through a wormhole.  Such plots follow straightforwardly from results computed using the $3+1$ formalism and we feel that such plots are a benefit of the $3+1$ formalism that are worth mentioning.

With the slicing condition $\alpha = \sqrt{A}$, our code is unable to evolve the system for too long after the formation of an apparent horizon.  The reason for this is that this slicing condition is unable to avoid the singularity that forms.  It is possible to use singularity avoiding slicing conditions that can continue the evolution further \cite{AlcubierreBook, BaumgarteBook}.  They accomplish this by decreasing $\alpha$ in regions where the singularity will form so that the evolution is slowed or ceases in this region, but continues elsewhere.  Since the formation of the apparent horizon is accompanied by the collapse of the wormhole throat, it might be possible to avoid the singularity by tying $\alpha$ to the metric field $C$, since $R = \sqrt{C}$.  Indeed, the slicing condition
\begin{equation} \label{C slicing}
\alpha(t,r) = \sqrt{A(t,r)} \, \frac{C(t,r)}{r_0^2+r^2}
\end{equation}
was shown to have strong singularity avoidance properties in \cite{Gonzalez:2008xk}.  We too have found that this slicing condition works well for the massless symmetric wormhole and we use it for the following results.

That regular matter leads to the collapse of the wormhole has led to the speculation that once a traveler passes through the wormhole, they will be unable to return.  It was shown in \cite{Shinkai:2002gv} that sending a second pulse of ghost matter after the first pulse of regular matter could keep the wormhole open.  We consider a different possibility.  We imagine sending a probe, made of regular matter, through the wormhole and ask if the probe can send a signal back through the wormhole.

\begin{figure}
\centering
\includegraphics[width=3.25in]{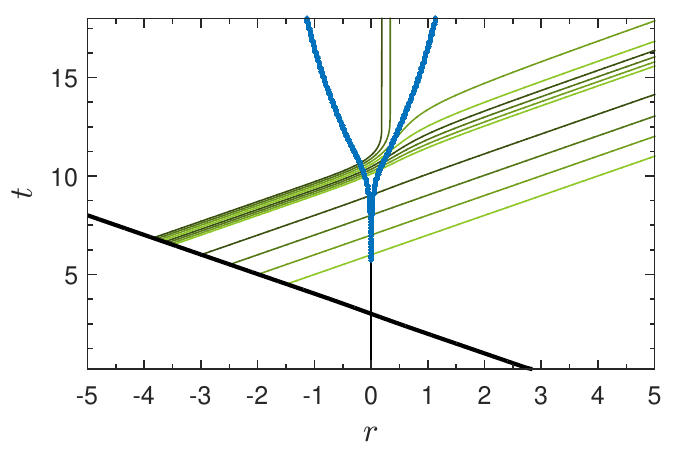}
\caption{The vertical black line at the bottom plots the position of the wormhole throat and the blue curve plots the apparent horizon.  These are the same quantities plotted in Fig.\ \ref{fig:massless regular R}(b), except that the slicing condition in Eq.\ (\ref{C slicing}) allows the evolution to continue further in time.  The thick black curve plots the position of the peak of the leftmost spike in Fig.\ \ref{fig:massless regular}, i.e.\ the position of regular matter as it travels through the wormhole.  The remaining green curves are null geodesics, representing the path of light signals sent from the matter field.  We can see that some of the light signals are able to travel back through the wormhole before it collapses.}
\label{fig:massless null}
\end{figure} 

\begin{figure*}
\centering
\includegraphics[width=7in]{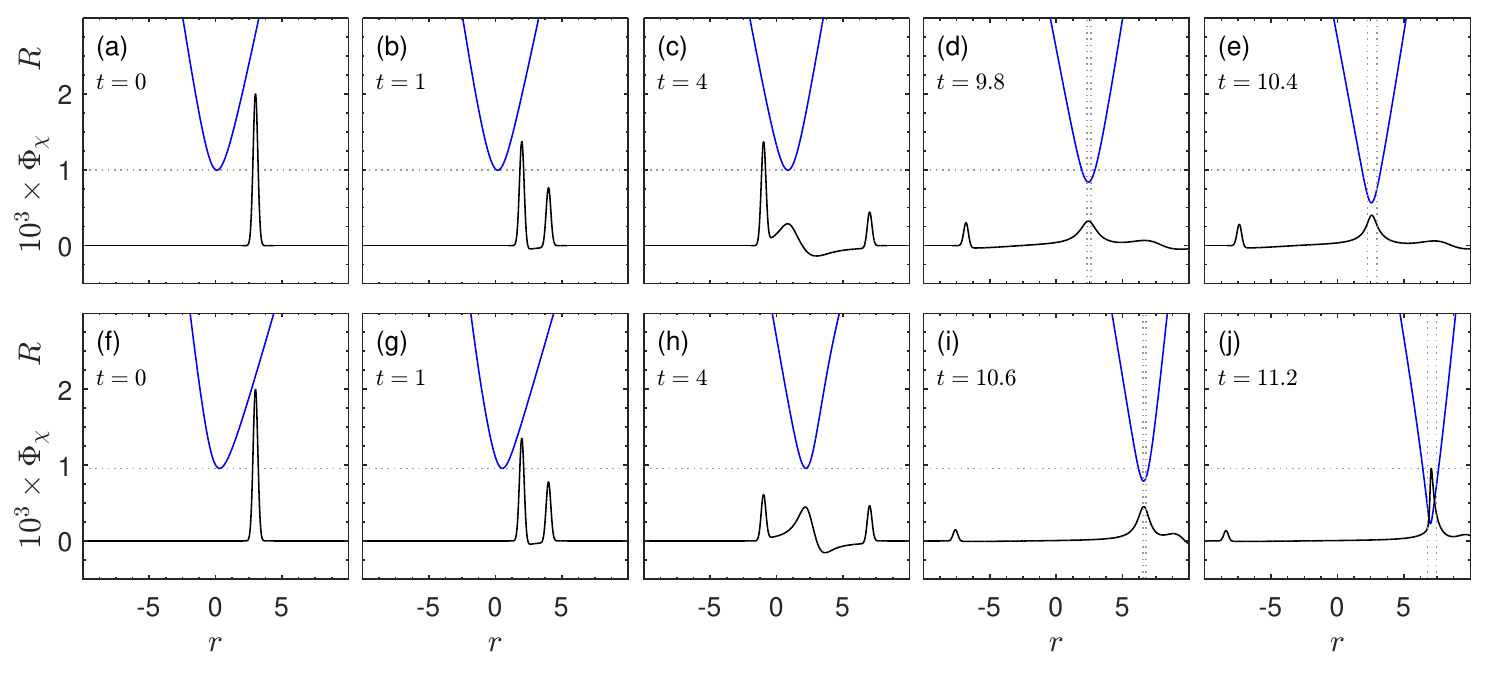}
\caption{Analogous to Fig.\ \ref{fig:massless regular}, except for massive asymmetric wormholes.  The wormhole in the top row is defined by $A_1 = 0.1$ and the wormhole in the bottom row is defined by $A_1 = 0.3$.}
\label{fig:massive regular}
\end{figure*} 

In Fig.\ \ref{fig:massless null}, we show an evolution using the same initial data as used in Fig.\ \ref{fig:massless regular}, but using the slicing condition in Eq.\ (\ref{C slicing}).  The vertical line at the bottom plots $r_\text{min}$, the position of the wormhole throat, and the blue curve plots the apparent horizon.  This is the same thing plotted in Fig.\ \ref{fig:massless regular R}(b), but the new slicing condition is able to evolve the system further in time.  The thick black curve plots the position of the regular matter traveling through the wormhole.  More specifically, it plots the position of the peak of the leftmost spike in Fig.\ \ref{fig:massless regular}.  The remaining (green) curves are null geodesics, computed using Eq.\ (\ref{null def}).  The null geodesics represent the path taken by a light signal emitted by our regular matter probe.  We can see that some of the light signals easily make it through the wormhole, some just barely make it, and a couple get caught inside the black hole.  Figure \ref{fig:massless null} shows us it is possible for the probe to send signals back through the wormhole before the wormhole collapses.

We end this section by commenting that in this wormhole geometry, since null geodesics are in some cases able to pass through the apparent horizon in Fig.\ \ref{fig:massless null}, the existence of an apparent horizon on a given time slice does not necessarily mean that there is an event horizon on the same time slice.  This phenomenon was seen in \cite{Gonzalez:2008xk}, whose analysis suggests that the apparent horizon asymptotically agrees with the event horizon.  We refer the reader to \cite{Gonzalez:2008xk} for a discussion.

\begin{figure}
\centering
\includegraphics[width=3.25in]{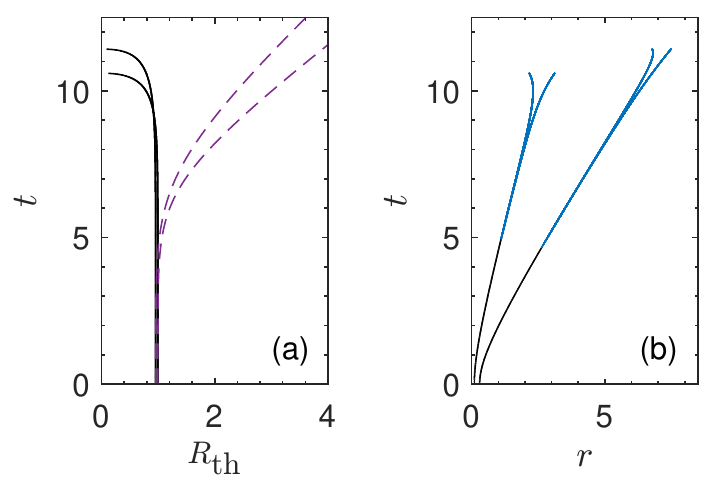}
\caption{Analogous to Fig.\ \ref{fig:massless regular R}, except for massive asymmetric wormholes.  (a) The solid black curves are for the same evolutions shown in Fig.\ \ref{fig:massive regular} and, from bottom to top, are for $A_1 = 0.1$ and $0.3$.  The dashed purple curves are for pulses of ghost matter and, from bottom to top, are for $A_1 = 0.1$ and $0.3$.  (b)  Both curves are for the same evolutions shown in Fig.\ \ref{fig:massive regular} and, from left to right, are for $A_1 = 0.1$ and $0.3$.}
\label{fig:massive regular R}
\end{figure}

\begin{figure*}
\centering
\includegraphics[width=7in]{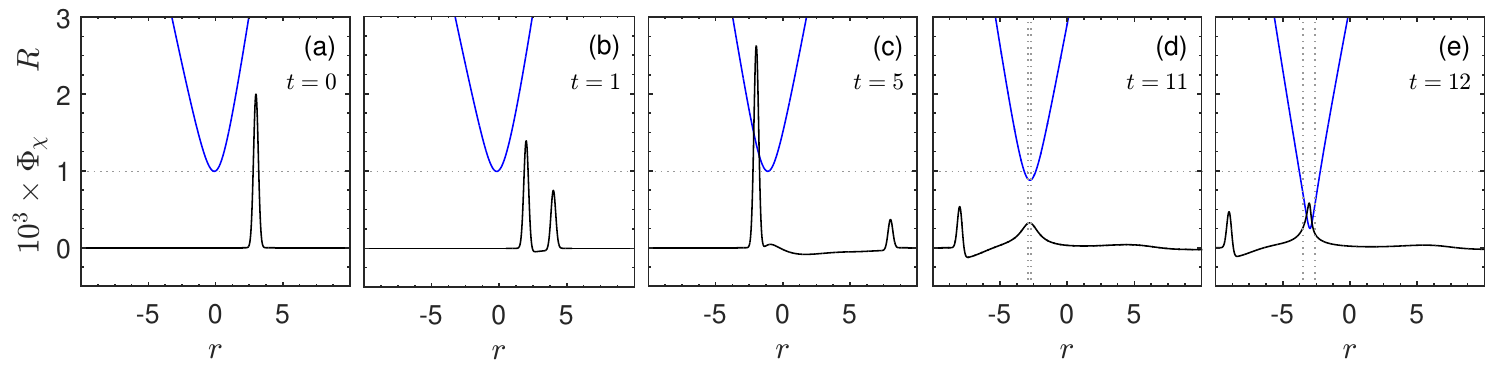}
\caption{Analogous to Figs.\ \ref{fig:massless regular} and \ref{fig:massive regular}, except for $A_1 = -0.1$.}
\label{fig:massive regular left}
\end{figure*}


\section{Massive asymmetric wormholes}
\label{sec:massive results}

In this section, we present results for the numerical evolution of massive asymmetric wormholes, which are defined by $A_1 \neq 0$.  Recall from Eq.\ (\ref{static rmin}) that the wormhole throat begins at $r_\text{min} = A_1$ (for $r_0=1$).  In the previous section, we found that the wormhole throat stays at $r = 0$.  Here, we find that massive wormholes move.  For positive $A_1$ the throat moves rightward, while for negative $A_1$ it moves leftward.  Beyond that, we find that the massive case is similar to the massless case: When a pulse of regular matter travels through the wormhole, the wormhole throat collapses and forms an apparent horizon.  For a pulse of ghost matter with negative amplitude it does the same, while for a pulse of ghost matter with positive amplitude it expands and does not form an apparent horizon.

In Fig.\ \ref{fig:massive regular}, we show results for the evolution of a pulse of regular scalar field, $\Phi_\chi$ (black curve), and the areal radius, $R$ (blue curve).  The top row is for $A_1 = 0.1$ and the bottom row is for $A_1 = 0.3$.  This figure is analogous to Fig.\ \ref{fig:massless regular} for the massless wormhole.  We can see that the wormhole throat travels rightward and that the pulse of regular matter leads to collapse and the formation of an apparent horizon.  We can also see that as we increase $A_1$, more matter gets stuck in the wormhole.  This can be seen by the larger amplitude for the piece that gets stuck and the smaller amplitude for the piece that travels through.

The same two evolutions shown in Fig.\ \ref{fig:massive regular} are shown in Fig.\ \ref{fig:massive regular R}, which is analogous to Fig.\ \ref{fig:massless regular R}.  In Fig.\ \ref{fig:massive regular R}(a), the two black curves, from bottom to top, are for $A_1 = 0.1$ and $0.3$, and display how the wormhole throat collapses.  In Fig.\ \ref{fig:massive regular R}(b), the two black curves display $r_\text{min}$ and, from left to right, are for $A_1 = 0.1$ and $0.3$.  The blue curves display apparent horizons.  For a pulse of ghost matter with positive amplitude, we find that the wormhole expands.  This is shown by the two dashed purple curves in Fig.\ \ref{fig:massless regular R}(a) which, from bottom to top, are for $A_1 = 0.1$ and $0.3$.  We have also simulated pulses of ghost matter with negative amplitude, which we do not show, and find that the wormhole throat collapses and forms an apparent horizon.

We now consider negative $A_1$.  We show in Fig.\ \ref{fig:massive regular left} a plot similar to Figs.\ \ref{fig:massless regular} and \ref{fig:massive regular} for a pulse of regular matter traveling through a wormhole with $A_1 = -0.1$.  We can see that the pulse is still able to make it through the wormhole, even though the wormhole is initially moving away from it.  We again find, for a pulse of regular matter, that the wormhole throat collapses.

\begin{figure}
\centering
\includegraphics[width=3.25in]{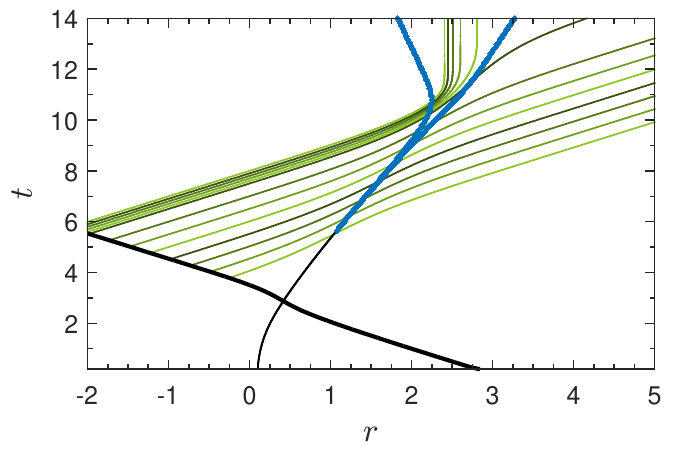}
\caption{Analogous to Fig.\ \ref{fig:massless null}, except using the same initial data used for the top row of Fig.\ \ref{fig:massive regular} and the left curve in Fig.\ \ref{fig:massive regular R}(b) and using the slicing condition in Eq.\ (\ref{exp C slicing}).}
\label{fig:massive null}
\end{figure}

Lastly, we consider sending a probe of regular matter through a massive asymmetric wormhole and ask if the probe can send a light signal back through the wormhole.  In the previous section, when answering this question for the massless wormhole, we used the singularity avoiding slicing condition in Eq.\ (\ref{C slicing}).  Unfortunately, we have not found that this slicing condition works well for the massive case.  Instead, we have had success with
\begin{equation} \label{exp C slicing}
\alpha(t,r) = \sqrt{A(t,r)} \left[ 1 - e^{-C(t,r)} \right],
\end{equation}
as long as $A_1$ is not too large.   In Fig.\ \ref{fig:massive null}, we show an evolution using the same initial data as the top row of Fig.\ \ref{fig:massive regular} and the left curve in Fig.\ \ref{fig:massive regular R}(b), but using the slicing condition in Eq.\ (\ref{exp C slicing}).  Figure \ref{fig:massive null} is analogous to Fig.\ \ref{fig:massless null} and shows that it is possible for a probe to send signals back through a massive asymmetric wormhole before it collapses.

\section{Conclusion}
\label{sec:conclusion}

Ellis-Bronnikov-Morris-Thorne wormholes are spherically symmetric analytical solutions for static wormholes constructed with a massless real ghost scalar field.  They are generically massive and asymmetric, with a massless symmetric wormhole as a special case.  As far as we are aware, they are the only wormholes to have been numerically evolved.  We revisited their numerical evolution, developing a new code based on the standard $3+1$ foliation of spacetime.  Although we focused on Ellis-Bronnikov-Morris-Thorne wormholes, we believe our code could be used more broadly for simulating wormhole geometries. 

For the massless wormhole, we confirmed that a pulse of regular scalar field causes the wormhole to collapse and form an apparent horizon.  For a pulse of ghost scalar field with negaitive amplitude we found the same, while for a pulse with positive amplitude we found that it causes the wormhole throat to expand and not form an apparent horizon.  We then showed that a pulse of regular matter can travel through the wormhole and send a light signal back before the wormhole collapses.  For a pulse of matter traveling through a massive wormhole, which had not previously been simulated, we again found that regular matter causes the wormhole to collapse and form an apparent horizon, as does a pulse of ghost matter with negative amplitude, while ghost matter with a positive amplitude causes the wormhole throat to expand.  We further showed that the wormhole moves, but that the pulse is still able to travel through it and that a pulse of regular matter can again send a light signal back through the wormhole before it collapses.


\acknowledgements

K.\ C.\ thanks the George Alden Trust for financial support.  B.\ F.\ thanks Shanahan for financial support.


\appendix

\section{Code tests}
\label{app:code tests}

\begin{figure*}
\centering
\includegraphics[width=6in]{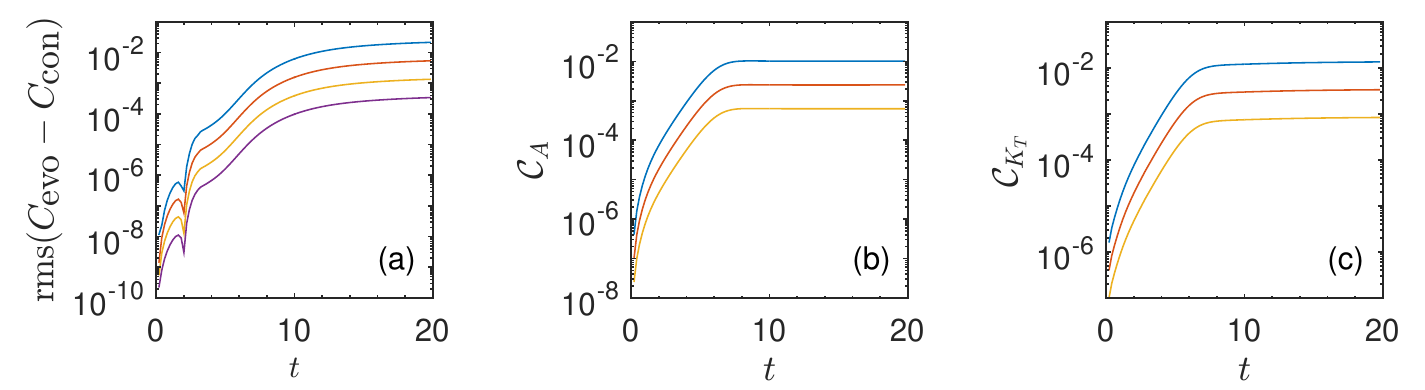}
\caption{Each curve uses the same initial data as used for the purple curve in Fig.\ \ref{fig:massless regular R}(a) for a ghost scalar field with positive amplitude traveling through the massless symmetric wormhole.  Each plot is made using grid spacings $\Delta r = 0.01$, 0.005, 0.0025, and 0.00125.  Neighboring curves in each plot drop by a factor of 4, indicating second order convergence.  See the Appendix for details.}
\label{fig:test 1}
\end{figure*}

\begin{figure*}
\centering
\includegraphics[width=6in]{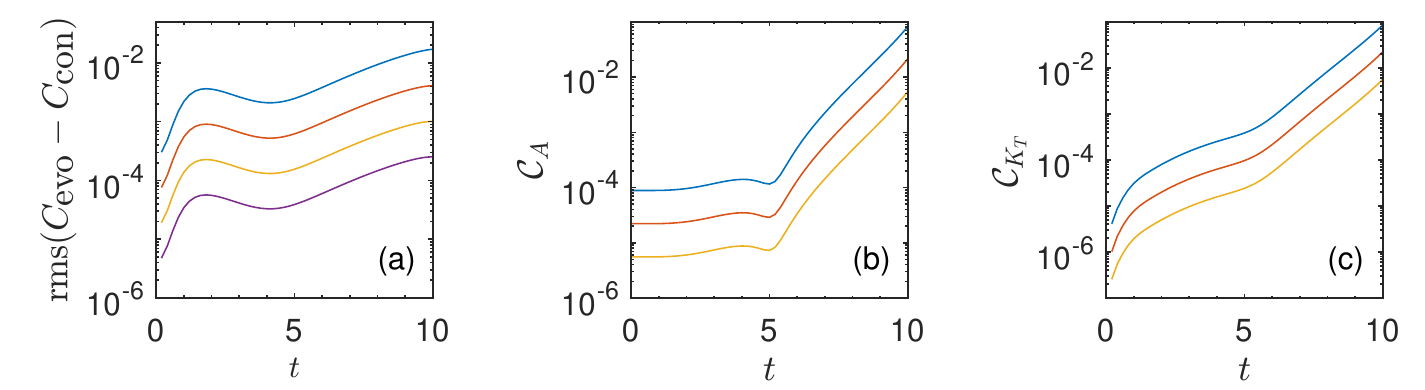}
\caption{Analogous to Fig.\ \ref{fig:test 1}, except for using the same initial data as used for the bottom row of Fig.\ \ref{fig:massive regular} for a regular scalar field traveling through a massive asymmetric wormhole.  Neighboring curves drop by a factor of 4, indicating second order convergence.}
\label{fig:test 2}
\end{figure*}

In this appendix, we present tests of our code and show that our code is second order convergent.  In Sec.\ \ref{sec:numerics}, we discussed how our code solves for $K_T$ using the momentum constraint and that we do not make use of the Hamiltonian constraint, both of which are in Eq.\ (\ref{con eqs}).  The Hamiltonian constraint is therefore available for code testing.  Specifically, since we solve an evolution equation for the metric field $C$, we can check to see if the result for $C$ satisfies the Hamiltonian constraint on each time slice, as it must.  We therefore compute \cite{AlcubierreBook}
\begin{equation} \label{rms C}
\text{rms}(C_\text{evo} - C_\text{con}),
\end{equation}
where $C_\text{evo}$ is the value of $C$ obtained from the evolution equation, $C_\text{con}$ is the value of $C$ obtained from the Hamiltonian constraint, and rms means to take the root-mean-square across the computational grid.  We compute this quantity with various grid spacings, $\Delta r$.  An indication of second order convergence is that this quantity drops by a factor of 4 when the grid spacing drops by a factor of 2 \cite{AlcubierreBook}.

We can also test our code without using a constraint equation as follows.  We define the convergence function \cite{AlcubierreBook}
\begin{equation}
\mathcal{C}_f^{\Delta r_1, \Delta r_2} \equiv \| f_\text{evo}^{\Delta r_1} - f_\text{evo}^{\Delta r_2} \|,
\end{equation}
where $f$ is an arbitrary field, $f_\text{evo}^{\Delta r_1}$ indicates the field as obtained from an evolution equation using grid spacing $\Delta r_1$, and $\|$ means to take the $L^2$ norm across the computational grid.  Note that $f_\text{evo}^{\Delta r_1}$ and $f_\text{evo}^{\Delta r_2}$ must be compared at a grid point and time step common to both simulations.  Using three different grid spacings, we can compute $\mathcal{C}_f^{\Delta r_1, \Delta r_2}$ and $\mathcal{C}_f^{\Delta r_2, \Delta r_3}$.  For $\Delta r_1/\Delta r_2 = 2$ and $\Delta r_2/\Delta r_3 = 2$, an indication of second order convergence is that $\mathcal{C}_f^{\Delta r_1, \Delta r_2} / \mathcal{C}_f^{\Delta r_2, \Delta r_3} = 4$ \cite{AlcubierreBook}.

We show some results in Fig.\ \ref{fig:test 1} using the same initial data used to compute the dashed purple curve in Fig.\ \ref{fig:massless regular R}(a) (i.e.\ a pulse of ghost matter with positive amplitude traveling through the massless symmetric wormhole).  The grid spacings used to make all three plots in Fig.\ \ref{fig:test 1} are $\Delta r = 0.01$, 0.005, 0.0025, and 0.00125.  In all three plots, neighboring curves drop by a factor of 4, indicating second order convergence.  We show in Fig.\ \ref{fig:test 2} an analogous plot but using the same initial data used to compute the bottom row of Fig.\ \ref{fig:massive regular} (i.e.\ a pulse of regular matter traveling through a massive asymmetric wormhole).  We again find that neighboring curves drop by a factor of 4, indicating second order convergence.  We also find that the curves rise at later times, which is not unexpected since a black hole forms and the code eventually loses numerical stability.  We find similar results for other fields and for other simulations.




%

\end{document}